\documentclass[reprint,superscriptaddress,amsmath,amssymb,aps,prb,showpacs]{revtex4-1}

\usepackage{gensymb}
\usepackage[]{graphicx,subfigure}
\usepackage{dcolumn}% Align table columns on decimal point
\usepackage{bm}% bold math
\usepackage{hyperref}
\usepackage{multirow}
\usepackage{tabularx}
\usepackage{siunitx}

\usepackage[utf8]{inputenc}

\begin{document}

\title{Characteristic diffuse scattering from distinct line roughnesses}

\author{A. Fern\'{a}ndez Herrero}
\affiliation{Physikalisch-Technische Bundesanstalt (PTB), Abbestr. 2-12, 10587 Berlin, Germany}

\author{M. Pflüger}
\affiliation{Physikalisch-Technische Bundesanstalt (PTB), Abbestr. 2-12, 10587 Berlin, Germany}

\author{J. Probst}
\affiliation{Helmholtz-Zentrum Berlin (HZB), Albert-Einstein-Str. 15, 12489 Berlin, Germany}

\author{F. Scholze}
\affiliation{Physikalisch-Technische Bundesanstalt (PTB), Abbestr. 2-12, 10587 Berlin, Germany}

\author{V. Soltwisch}
\affiliation{Physikalisch-Technische Bundesanstalt (PTB), Abbestr. 2-12, 10587 Berlin, Germany}

\date{\today}

\begin{abstract}

Lamellar gratings are widely used diffractive optical elements; gratings etched into Si can be used as structural or prototypes of structural elements in integrated electronic circuits. For the control of the lithographic manufacturing process, a rapid in-line characterization of nanostructures is indispensable. Numerous studies on the determination of regular geometry parameters of lamellar gratings from optical and Extreme Ultraviolet (EUV) scattering highlight the impact of roughness on the optical performance as well as on the reconstruction of these structures. Thus, a set of nine lamellar Si-gratings with a well-defined line edge roughness or line width roughness was designed. The investigation of these structures using EUV small angle scattering reveals a strong correlation between the type of line roughness and the angular scattering distribution. These distinct scatter patterns open new paths for the unequivocal characterization of such structures by EUV scatterometry.

\end{abstract}

\maketitle
\section{introduction}
Lithographically manufactured nanostructures play an important role as structural elements of integrated electronic circuits. With shrinking structure sizes, the impact of the roughness has gained more influence on their performance.
The demand for better manufactured nanostructures has motivated the development of methods and techniques for structure and roughness analysis.
For high resolution surface analysis, scanning techniques have been widely used. These methods have the advantage of measuring in real space and they have already been used for the characterization of the roughness in nanopatterned structures ~\cite{fouchier_atomic_2013,patsis_roughness_2003}.
Despite the advantages of these direct techniques, long measuring times are needed if one is to obtain relevant statistical information.
However, for the control of the lithographic manufacturing process, a rapid in-line and alteration-free characterization of such structures is indispensable ~\cite{bunday_hvm_2016}. 
Indirect optical methods are non-destructive techniques with a much lower acquisition time. Besides the well-established optical methods, also X-ray methods are investigated for future metrology solutions ~\cite{Lemaillet_intercomparison_2013}. The measurements can be performed in transmission geometry, such as in Critical Dimension Small-Angle X-ray Scattering (cd-SAXS) ~\cite{thiel_advances_2011} or in reflection geometry, where the enhancement of the surface signal can be obtained by illuminating the sample at a grazing incident angle, close to the critical angle of reflection.

Grazing incidence small angle scattering is a well-established ~\cite{muller-buschbaum_grazing_2003,muller-buschbaum_basic_2009,hofmann_grazing_2009}, rapid and non-destructive technique used for the characterization of nanopatterned surfaces.
The scattering pattern provides statistical information about the illuminated area. Nevertheless, it is an indirect technique which requires a non-straightforward data interpretation ~\cite{renaud_probing_2009}, i.e. the structures and their uncertainties must be reconstructed from the scattered intensities. In this regard, reasonable understanding has been achieved when laterally periodic samples are investigated ~\cite{baumbach_grazing_1999,yan_intersection_2007,wernecke_direct_2012}.
Significant diffuse scattering, caused by rough lamellar gratings, has been constantly reported ~\cite{rueda_grazing-incidence_2012,suh_characterization_2016,soltwisch_correlated_2016}. Recently, also all dominant diffuse scatter contributions could be attributed to basic scattering processes ~\cite{soltwisch_correlated_2016} similar to scattering in multilayer systems. Roughness in multilayer systems is usually understood as surface or interface roughness, which has been widely studied by X-ray scattering and reflectivity ~\cite{mikulik_x-ray_1999,haase_multiparameter_2016}. In latterally nanopatterned structures the equivalent to interface roughness is the unevenness of the sidewalls. In an idealized figure, in a two-dimensional representation of the nanostructured surface, this can be described by the edge roughness, i.e. the deviation of the actual edge contour from a straight line. For line-and-space structures, it is usually classified into two types: line edge roughness (LER) where the line centre position varies along the line and line width roughness (LWR), where the width of the line changes along the line. Although there is no pure LER nor LWR in real structures, this distinction has been made for simplification and the separate study of the impact of the roughness in several analyses ~\cite{wang_characterization_2007,kato_effect_2010}. These calculations use Fourier optics and a binary grating, and conclude that a Debye-Waller-like factor can describe the impact of the roughness on the diffraction efficiency. Following these results one could assume that for the description of real samples the superposition of periodic functions might be sufficient. But this assumption has not been corroborated yet or exploited to characterize the roughness-induced scatter of real samples.

Considering those previous reports, we have designed a set of nine gratings, comprising LWR and LER with different distributions to be investigated by Extreme Ultraviolet (EUV) scatterometry. EUV small angle scattering exploits the high sensitivity of grazing incidence techniques while reducing the beam footprint, due to the larger incidence angles; which allows the investigation of smaller samples. The study presented here aims to provide a better understanding of scattering caused by line roughness. 
The samples were illuminated at several grazing incidence angles and the specular reflection and the discrete diffraction orders as well as the diffuse scattering distributions were recorded. The distinct distribution of the scattering patterns opens new paths for the unequivocal characterization of such structures by EUV scatterometry.
\section{Experimental Details}
\subsection{EUV-Small Angle Scattering}

The experiments were conducted at the soft X-ray beamline ~\cite{scholze_high-accuracy_2001} of the Physikalisch-Technische Bundesanstalt (PTB), which covers the photon energy range from $50$ eV to $1700$ eV, at the electron storage ring BESSY II.

\begin{figure}[htbp]
	\centering
    \includegraphics[width=0.40\textwidth]{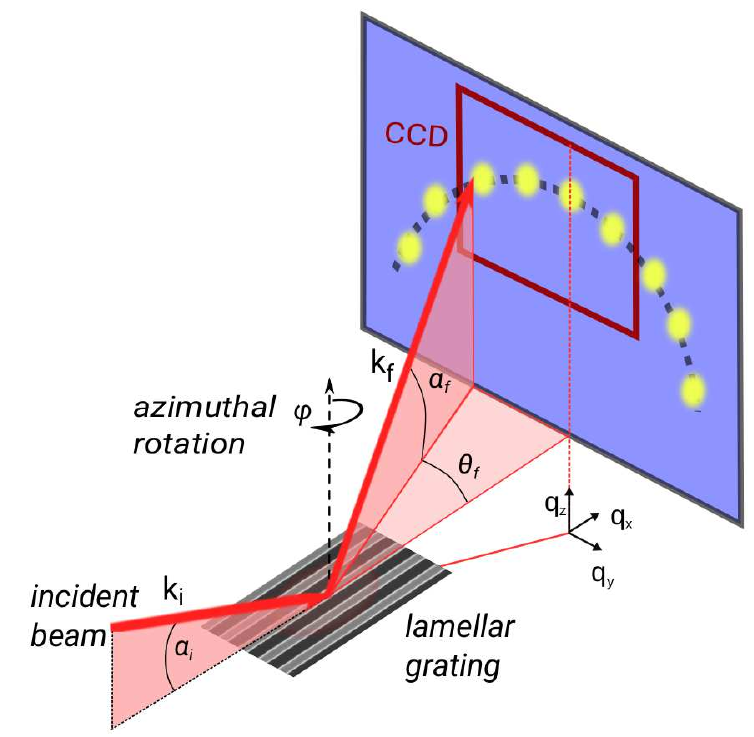}
	\caption{Experimental set-up. The red frame shows the area covered by our CCD. }
	\label{figure_sketch}
\end{figure}

The experimental set-up is illustrated in Fig.~\ref{figure_sketch}. A monochromatic X-ray beam with a wavevector $\vec{k}_i$ impinges on the sample surface at an incidence angle $\alpha_i$. The elastically scattered wavevector $\vec{k}_f$ propagates along \textbf{the direction with} exit angle $\alpha_f$ and the azimuthal angle $\theta_f$. The sample is illuminated in a conical diffraction mounting with the incidence plane parallel to the grating lines, $\varphi=0$, and placed in a goniometer, which allows us to rotate and move the sample,  inside a vacuum chamber. The detector is a $2048 \times 2048$ pixel Andor CCD camera with a pixel size of $13.5$ $\mu$m, which is placed at $14^\circ$ to the incoming beam and at $740$ mm off the sample, covering a field of view of approximately $2^\circ \times 2^\circ$ with the specular reflection of $\alpha_i = 7 ^\circ$ centered at the CCD. 
The orders of diffraction are given by the intersection of the Ewald sphere of elastic scattering with the reciprocal lattice. The coordinates in reciprocal space correspond to the momentum transfer:

\begin{equation}
\begin{pmatrix}

q_x \\ %= \frac{2\pi}{\lambda}(\cos(\theta_f) \cos(\alpha_f) - cos(\alpha_i)),\\
q_y \\ %= \frac{2\pi}{\lambda}(\sin(\theta_f) \cos(\alpha_f)),\\
q_z   %\frac{2\pi}{\lambda}(\sin(\alpha_f) + \sin(\alpha_i)).\\

\end{pmatrix}
= \frac{2\pi}{\lambda}
\begin{pmatrix}
 \cos(\theta_f) \cos(\alpha_f) - cos(\alpha_i)\\
 \sin(\theta_f) \cos(\alpha_f)\\
\sin(\alpha_f) + \sin(\alpha_i)
\end{pmatrix}
\label{eq_q}
\end{equation}
\noindent 
If the projection of the incidence plane is parallel to the grating lines $\varphi = 0$; the diffraction orders describe a semicircle in the detector plane~\cite{mikulik_coplanar_2001,yan_intersection_2007} with its center at the intersection of the sample horizon and the specular reflection plane. Therefore, the azimuthal angle $\varphi$ was aligned such that this condition was met, with the elevation angle from the sample horizon of the respective positive and negative diffraction orders being equal. The achieved angular uncertainty in $\varphi$ was $0.02 ^\circ$, which is sufficient for this experiment.

The experiment is analogous to the well-known Grazing Incidence Small Angle X-ray Scattering (GISAXS) technique but working with lower photon energy which allows to use a larger incidence angle, significantly reducing the beam footprint on the sample. 
 
\subsection{Sample Design}

Nine Si-gratings with different types of roughness were prepared by e-beam lithography and reactive ion etching at the Helmholtz-Zentrum Berlin. Each grating has a size of \SI{4}{mm} by \SI{0.51}{mm}, with the lines parallel to the long direction. The first grating is a reference grating with no artificial roughness added. The other eight gratings were designed to accomplish a well-defined line roughness, introducing perturbations to the ideal grating. The parameters of the gratings, such as pitch and linewidth, were chosen to be compatible with the roughness amplitudes. The width of the trenches between the lines influences the etching rate into the silicon during the reactive ion etching process such that for too narrow trenches the etch depth is reduced. Therefore, the pitch and linewidth, and the perturbations introduced were chosen to keep the trench width above a threshold of \SI{65}{\nm} to assure constant etch depth along the gratings. Thus, the pitch is \SI{150}{\nano\metre}, the nominal linewidth \SI{65}{\nano\metre} and the nominal etch depth, i.e. line height \SI{120}{\nano\metre}.

\begin{figure}[h!]
\centering
    \includegraphics[width=0.48\textwidth]{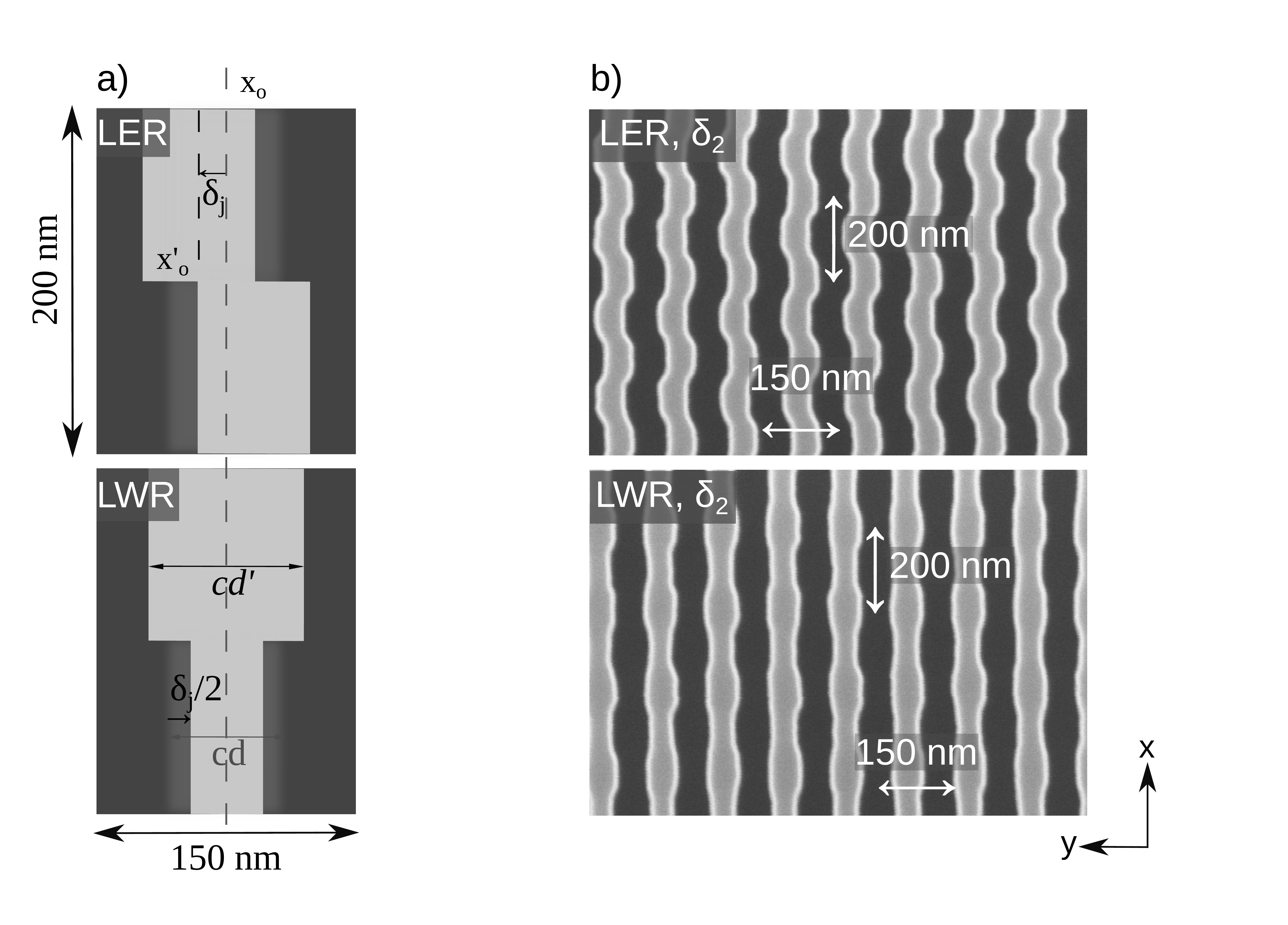}
\caption{Periodic roughness: a) Basis cell for the construction of the periodic LER (up) and LWR (bottom). b) SEM images of a sample with periodic LER (up) and LWR (bottom) with pitch of \SI{150}{\nano\metre} and a period of the roughness of \SI{200}{\nano\metre}.} 
\label{periodic_sem_sketch}
\end{figure}

\begin{figure}[htbp]
	\centering
    \includegraphics[width=0.48\textwidth]{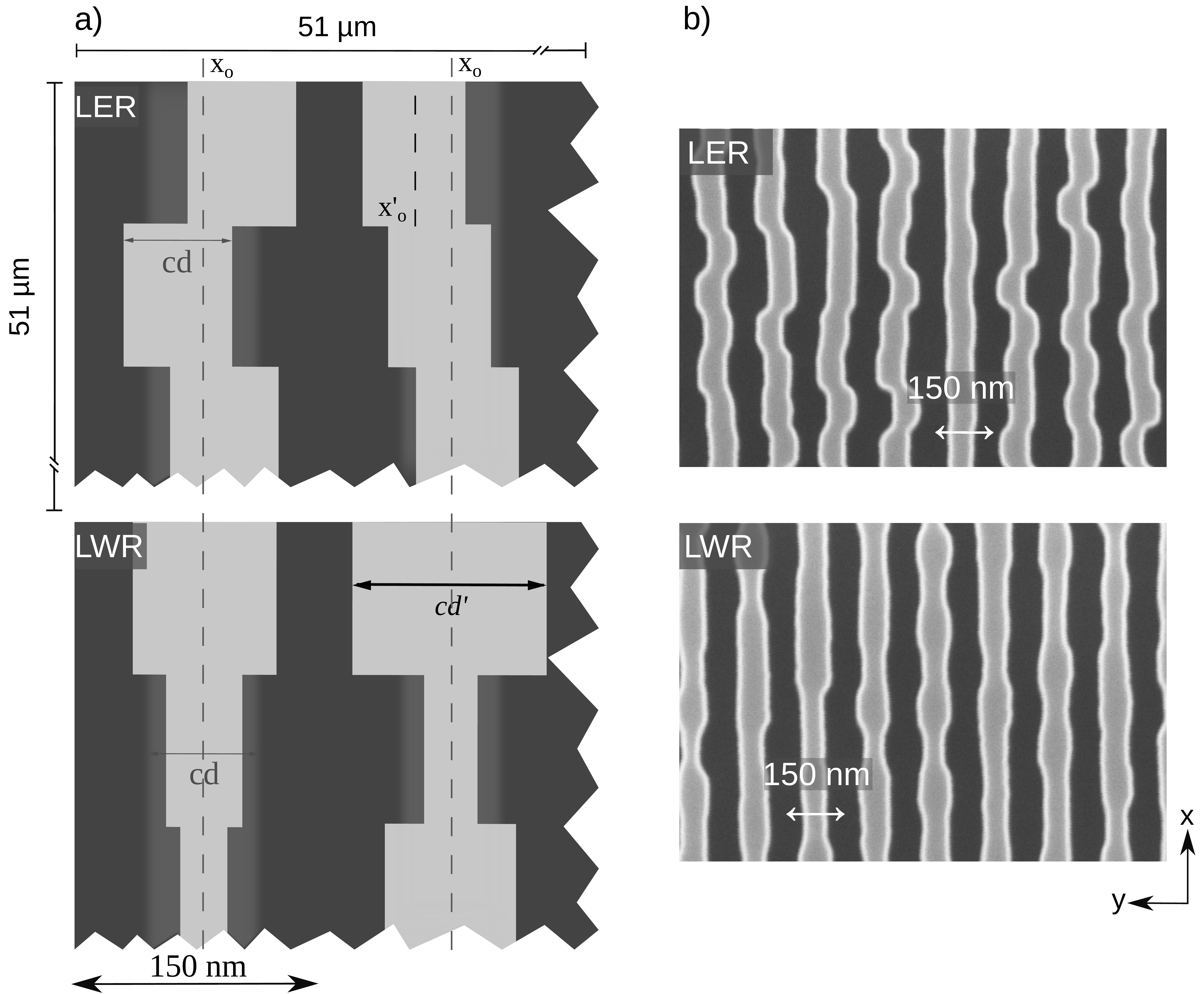}
\caption{Stochastic roughness: a) Design of a basis cell of \SI{51}{\um}, where each line follows a uniform distribution. b) SEM images of a sample with LER (top) and LWR (bottom).}
\label{stochastic_sem_sketch}
\end{figure}

Following the previous studies on the impact of the line roughness, we have designed four samples with a periodic roughness distribution. However, in order to study the effect of the roughness in real samples we have completed the set with another four gratings with a stochastic roughness distribution. Two samples of each distribution correspond to a different type of line roughness: line edge roughness or line width roughness. For the LER, the size of the pitch and of the linewidth are maintained constant, but the line centre position along the line is changed. On the other hand, for the LWR the centre position is kept constant, and also the pitch, while the width of the line is varied along the line. 
For the reference grating, each line can be understood as a chain of juxtaposed boxes \SI{100}{\nano\metre} long and  \SI{65}{\nano\metre} wide, centred at $x_0$ in a nominal pitch of \SI{150}{\nano\metre}; for the other eight gratings a perturbation is introduced to each of the boxes (see Fig.~\ref{periodic_sem_sketch} a) or Fig.~\ref{stochastic_sem_sketch} a)).\\

For the periodic roughness, we consider a basis cell composed of two adjacent boxes centred in the pitch, i.e.\ a size of \SI{200}{\nm} by \SI{150}{\nm}. A positive or negative perturbation, $\delta$, is introduced alternatively to the centre position of each box, in the LER frame, or to the width of the box, for the samples with LWR (see Fig.~\ref{periodic_sem_sketch} (a)). This basis cell is repeated until the sample area of \SI{0.51}{\mm} by \SI{4}{\mm} is completed. Thus there are two periodic dimensions: the intrinsic periodicity of the grating, i.e. the pitch, $p$, in the \textit{y}-direction and an artificial periodicity caused by the periodic roughness distribution, $p_r$, \SI{200}{\nano\metre} long along the line (in \textit{x}-direction) and identical lines are placed next to each other (see Fig.~\ref{periodic_sem_sketch} (b)). For each type of line roughness, two samples, each of them with a different perturbation amplitude $\delta_j$, were prepared, $\delta_1 = \SI{5}{\nm}$ and $\delta_2 = \SI{10}{\nm}$. 

Thus, adjacent blocks in the LER structure have a constant centre displacement and adjacent blocks of the LWR structure have a constant difference in width of 2$\delta_j$.
\\

For the stochastic roughness a squared basis cell of \SI{51}{\um} side length was used. The size of the basis cell was chosen as a compromise to, on the one hand, limit the amount of data required for the e-beam writing process and on the other hand, to be large enough that, when repeated to fill the \SI{4}{\mm} length of the grating, no observable periodicity is expected. Every \SI{150}{\nm} a new line of juxtaposed boxes is built, each of these lines follows a uniform discrete distribution. As was done for the periodic roughness, two samples with different amplitudes are designed with LER and another two with LWR. For each type of line roughness, one samples has a maximum perturbation amplitude of $\SI{10}{\nm}$ and the other of \SI{20}{\nm}. The e-beam writer's resolution was set to be \SI{1}{\nm} because smaller values would have increased the writing time drastically, which means all line boundaries are located on this discrete \SI{1}{\nm} grid. Therefore, the distribution of the actual perturbation amplitude is in discrete steps of \SI{1}{\nm} for the LER samples but of \SI{2}{\nm} for the LWR samples, as a perturbation in the latter type involves the displacement of two boundaries in opposite directions.

In these samples, there is no constant displacement or width difference between juxtaposed boxes but rather a stepwise interval from a \SI{0}{\nm} difference to a maximum of \SI{20}{\nm} or \SI{40}{\nm}, respectively.

\section{Results}

The scatter distributions presented here were measured at a wavelength of \SI{1.24}{\nm}, or a photon energy of 1 keV, and three different grazing angles. The CCD camera was positioned at $\ang{14}$ to the direct beam. The first incidence angle, $\alpha_i=\ang{7}$ was chosen to have specular reflection centred at the CCD camera. For further images, the sample was rocked until the orders of diffraction were out of the CCD image area ($\alpha_i=\ang{6.45},\ang{5.54}$) to enable long exposures of the diffuse scatter distributions without saturating the CCD camera. 

\begin{figure}[htbp]
	\centering
    \includegraphics[width=0.35\textwidth]{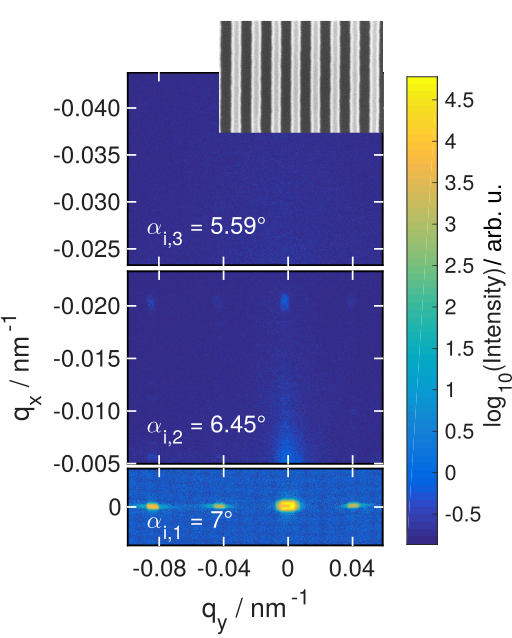}
\caption{2D scattering pattern from the reference grating. No significant diffuse scattering is observable.}
\label{reference}
\end{figure}

\begin{figure*}[htbp]
	\centering    \includegraphics[width=0.75\textwidth]{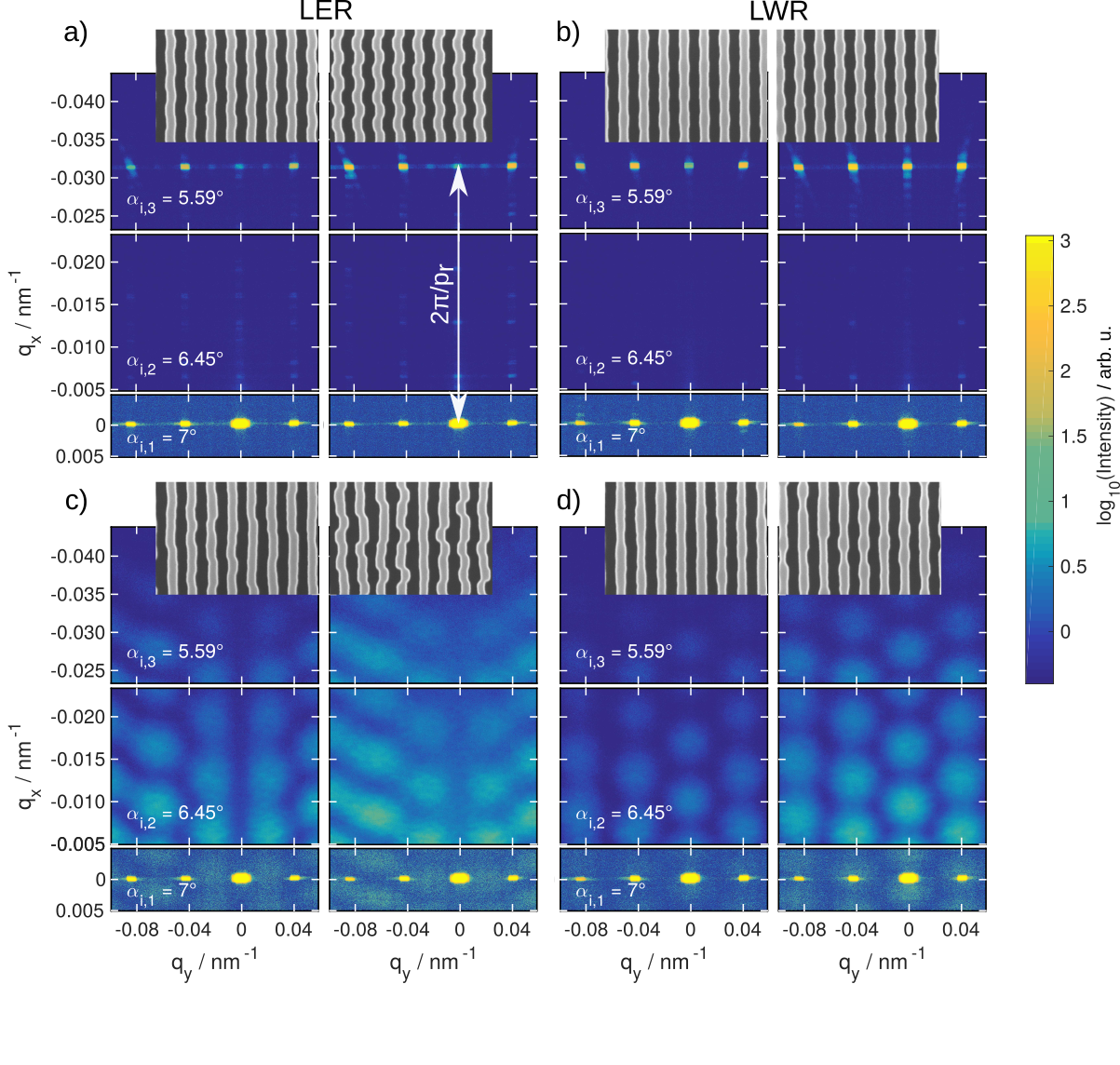} \\    
\caption{2D scattering pattern from the eight rough lamellar gratings. From the sample with periodic roughness: a) LER and b) LWR. The attenuation of the zero satellite order ($\alpha_i= \SI{5.54}{\deg}$, $q_y=0$) happens for samples with LER. And from the stochastic LER (c)) and LWR (d)), where the distinct scattering distribution is clearly visible for each type of line roughness. 
}
\label{figure_diffuse_regular}
\end{figure*}

The data are represented in the $(q_x, q_y)$ momentum space; the discrete orders of diffraction for the regular grating are found at $q_x = 0$ (see Fig.~\ref{reference}). Due to the larger footprint of the beam compared to the grating size, we have the contribution of the specular reflection coming from the substrate, which is superposed to the zero order of diffraction, see the larger spot size of the zero order shown in Fig.\ref{reference}. The distance between the orders of diffraction in $q_y$ corresponds to the pitch size, $\Delta q_y=\frac{2\pi}{p}$. Then the sample was rocked and measured  for the other angles of incidence. We acquired these images with a longer exposure time. All rough samples show intense contributions which are not observable at the reference grating due to the high quality of the structure (compare Fig. \ref{reference} and Fig. \ref{figure_diffuse_regular}). Note that just the samples with a stochastic distribution present diffuse scattering patterns.
\subsection{Periodic Line Roughness}

The four samples with a periodic roughness have a two-dimensional periodicity: one is the pitch of the grating, $p$, along the \textit{y}-axis and the second one is the roughness periodicity, $p_r$, in the \textit{x}-direction. Due to this particularity, satellite orders are observable at frequencies $q_x= n \frac{2\pi}{p_r}$ with integer n, see Fig.~\ref{figure_diffuse_regular} a) and b). These periodicities dominate the intensity distribution of the out-of-plane scattering pattern. Thus, for the samples with a periodic roughness no diffuse scatter is observed but the light scattering is fully governed by the two-directional periodicity of the sample. 
This case of periodic roughness was already studied using  Fourier optics. Several authors have focused their work on the behaviour and the impact of the roughness on the diffraction orders, for instance~\citeauthor{kato_analytical_2012} and~\citeauthor{wang_characterization_2007}. Both studies consider a binary model and distinguish between LER and LWR. Their proposed models and their applicability are discussed here.

The distribution of the satellite orders is qualitatively similar to the results obtained by ~\citeauthor{kato_analytical_2012}. Here the authors showed that for a grating with a sinusoidal line-width or -position variation, the intensity of the satellite orders is given by Bessel functions of the first kind. For LER the intensity of the satellite peaks at $q_y= 0$ diminishes while for LWR only the first order satellite peak at $q_y= 0$ exists. This is clearly confirmed by our measurements, where the satellite peak at $q_y=0$ for samples with LER is only very weak as compared to the LWR samples, see Fig.~\ref{figure_diffuse_regular} b) and c) for $\alpha_{i,3} = 5.59^\circ$. The remaining low intensity for the LER sample can be explained by the structures having a small amount of LWR also present in the LER structures.

However, we observe that the two-dimensional model considered by~\citeauthor{kato_analytical_2012} is not sufficient for the characterization of the roughness amplitudes. In the same way, the model described by~\citeauthor{wang_characterization_2007}, which uses a unit cell with a two-dimensional lattice formed by the stack of two blocks, and carried out in the frame of transmission cd-SAXS, is not applicable to a grazing incident measurement. By considering a binary grating, the influence of the height of the structure on the intensity distribution, i.e. the possible effect in the $q_z$ component of the scattering vector, is underestimated. In the frame of cd-SAXS, in the study of~\citeauthor{wang_characterization_2007} this effect is totally disregarded as the $q_z$ changes are not observable in these measurements. In the same way, other studies performed in transmission cd-SAXS, for instance~\citeauthor{freychet_study_2016}, can not be directly compared to the results discussed here.

Therefore, we can conclude that the off-specular scattering intensity distribution from samples with periodic roughness is dominated by the two-dimensional pitch. However, the existing studies do not consider the corresponding changes in the $q_z$ component of the scattering vector which strongly influences the intensity of the satellite orders. This effect is clearly visible in the diffuse scattering patterns given by the samples with stochastic roughness in the next section.

\subsection{Stochastic Line Roughness}

In contrast to the patterns from samples with periodic roughness, samples with stochastic line roughness just have the periodicity given by the pitch $p$, and therefore no satellite peaks are observed but a purely diffuse scattering pattern. In multilayer systems, the resonant diffuse scattering (RDS) is well known. It appears due to the correlation of the roughness of the interfaces ~\cite{kaganer_bragg_1995,holy_nonspecular_1994}. For lamellar gratings these effects were already reported ~\cite{soltwisch_correlated_2016,hlaing_nanoimprint-induced_2011} in the form of palm-like diffuse scattering sheets, caused by interference within the \textit{effective layer} of the grating.

In our case, depending on the type of line roughness, LER or LWR, the diffuse scattering pattern shows a different angular intensity distribution (see Fig.~\ref{figure_diffuse_regular} c) (LER) and d) (LWR)). A combination of the LER and LWR contributions corresponds to the palm-like pattern usually observed when real samples are investigated. 

Here each type of roughness leads to a characteristic diffuse scattering pattern. In line with the observation for the periodic samples, no intensity is observable for the off-specular scattering of the LER structures at $q_y=0$. The maxima of the diffuse scatter are shifted by a half-period with respect to the discrete diffraction orders. This is particularly notable because the diffraction intensities for the quasi-periodic samples are, naturally, aligned with the discrete diffraction orders of the regular grating. It is therefore not possible to obtain the scattering pattern of the non-regular LER sample from a superposition of quasi-peroidic solutions. In contrast, for LWR diffuse scattering along $q_y=0$ is observed and the diffuse scatter distributions are in phase with the discrete diffraction orders. This difference allows distinguishing between samples with LER and LWR. This difference on the behaviour of the constructive interference for samples with LER and LWR also implies that the rigorous calculations, applicable for periodic roughness distributions, cannot be applied to a sample with a stochastic roughness distribution.

\begin{figure}[htbp]

	\centering
    \includegraphics[width=0.45\textwidth]{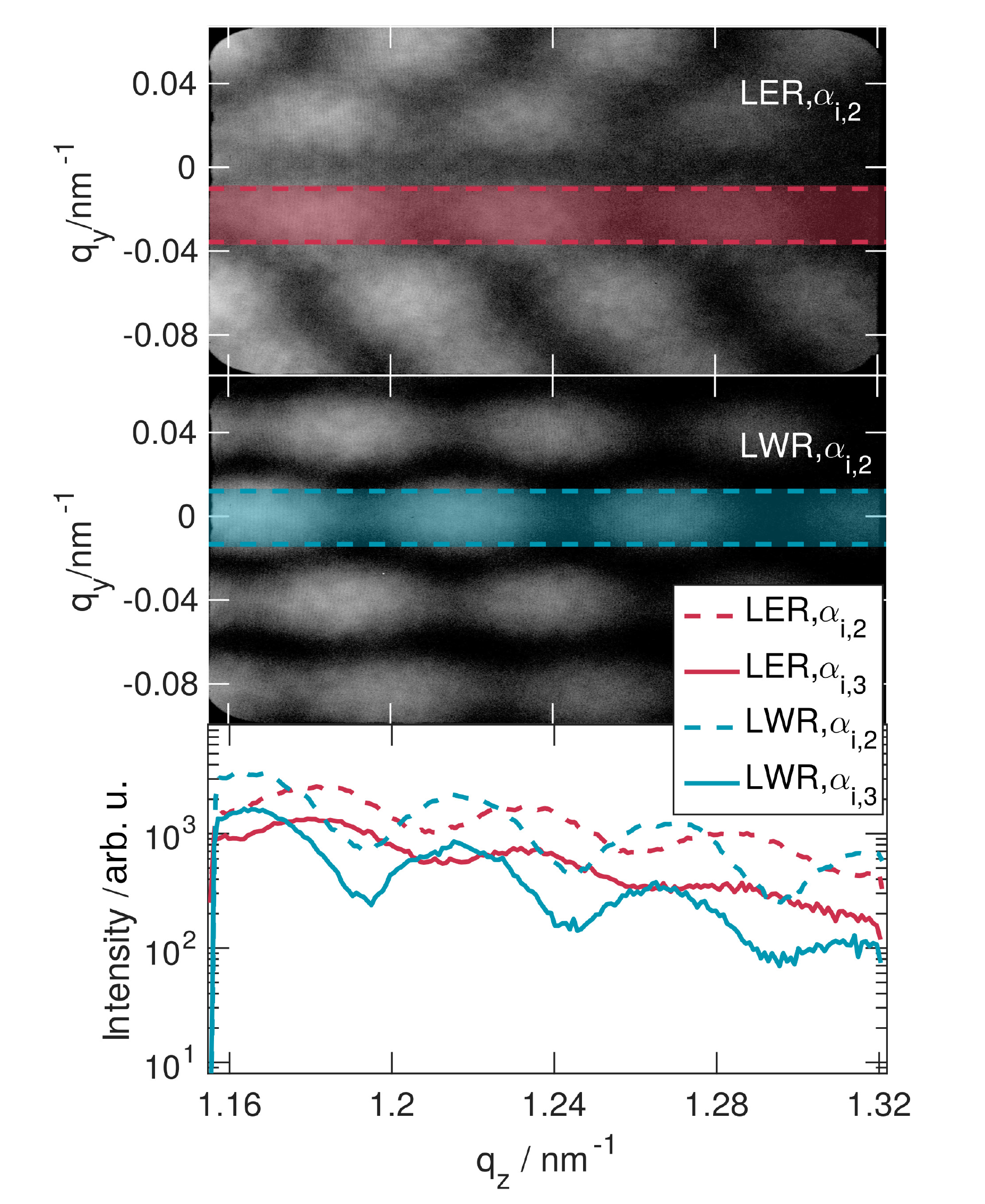}
\caption{The curves in the lower panel show the integration of the signal from the $(q_y,q_z)$ maps (shown above) for one amplitude of the LER (LWR) and the two incident angles where the regular orders are out of field ($\alpha_{i,2}$ and $\alpha_{i,3}$). For the sample with LER (red) the integration was done around $q_y = - \pi/p$, and for the LWR (blue) around $q_y = 0$. In the $(q_y,q_z)$ maps, the integration areas (red or blue, respectively) are shaded.}
\label{cut_plot}
\end{figure}

Likewise the influence of the amplitude of the roughness does not influence the angular distribution of the diffuse scattering, just the intensity of the contributions. For a better illustration we also projected the scatter intensities in the $(q_y,q_z)$ plane. Fig.~\ref{cut_plot} shows a linear cuts of these projections along $q_z$. The observed modulation is well in phase for both roughness amplitudes in both cases, LER and LWR. We conclude that the palm like modulation is caused by the above mentioned interference effects in the grating \textit{effective layer}, i.e. a $q_z$ effect which is naturally not covered by the previous studies based on Fourier optics ~\cite{kato_analytical_2012}. 

\section{Conclusion}

We investigated a set of nine lamellar Si-gratings which comprises samples with different types of line roughness (LER and LWR), roughness distributions (periodic and stochastic) and amplitudes. 
The periodic and stochastic roughness samples present a different off-specular scatter distribution. Periodic roughness leads to a distribution of satellite orders given by the periodicity of the roughness and the periodicity of the original structure, the pitch. 
On the other hand, for samples with stochastic roughness a pure diffuse scattering pattern is observable.

We have observed a correlation between the type of roughness and the scatter distribution. The intensity of the satellite peaks from samples with periodic roughness provides information on the type of the roughness encountered. The satellite peaks, in $q_x \neq 0$, at $q_y=0$ are suppressed for samples with LER in contrast to samples with a predominant LWR. This finding confirms the previous rigorous calculations where it was stated that the samples with a predominant LER lead to the extinction of the zero-order satellites. However, the existing models do not take into account the $q_z$ effect  which is explored in our geometry.
For samples with stochastic roughness the distribution of the diffuse scattering strongly depends on the type of roughness. For the LER, no non-specular intensity is observable for $q_y=0$ and the diffuse scattering is phase shifted with respect to the regular diffraction orders of the periodic sample. For the LWR, non-specular intensity is observed at $q_y=0$ and the diffuse scattering is in phase with  the regular diffraction orders of the periodic sample. This fact questions previous analytical studies which proposed to use a superposition of solutions for periodic roughness to describe the behaviour of a real sample. Our finding for LER structures, in particular, shows that this is not feasible because of the phase shift in the diffuse scatter pattern. We could show that the diffuse scatter is periodic in $q_z$. The interpretation of the diffuse scattering distributions from these samples, where the parameters are known, opens new perspectives to the characterization of the roughness using EUV scatterometry.  
\bibliography{main.bbl}

\end{document}